\documentclass[aps,prl,amsmath,amssymb,amsfonts,twocolumn,showpacs,superscriptaddress,groupedaddress]{revtex4}
\usepackage{amsthm}
\usepackage{epsfig}
\usepackage{graphicx}
\usepackage{subfigure}
\usepackage{dcolumn}
\usepackage{bm}
\usepackage{color}
\usepackage[normalem]{ulem}

\newcommand{\del}{\partial}

\newcommand{\sign}{\operatorname{sign}}

\newtheorem*{theorem}{Theorem}

\begin{document}

\title{Quantum fields and entanglement on a curved lightfront}

\author{Illan Halpern}
\email{illan@berkeley.edu}
\author{Yasha Neiman}
\email{yashula@gmail.com}
\affiliation{Perimeter Institute for Theoretical Physics, 31 Caroline St N, Waterloo, ON, N2L 2Y5, Canada}

\date{\today}

\begin{abstract}
We consider field quantization on an arbitrary null hypersurface in curved spacetime. We discuss the de Sitter horizon as the simplest example, relating the horizon quantization to the standard Fock space in the cosmological patch. We stress the universality of null-hypersurface kinematics, using it to generalize the Unruh effect to vacuum or thermal states with respect to null ``time translations'' on arbitrary (e.g. non-stationary) horizons. Finally, we consider a general pure state on a null hypersurface, which is divided into past and future halves, as when a bifurcation surface divides an event horizon. We present a closed-form recipe for reducing such a pure state into a mixed state on each half-hypersurface. This provides a framework for describing entanglement between spacetime regions directly in terms of their causal horizons. To illustrate our state-reduction recipe, we use it to derive the Unruh effect.
\end{abstract}


\maketitle


\section{Introduction and summary}

\enlargethispage{.5\baselineskip}

Lightfront quantization \cite{Leutwyler:1970wn,Brodsky:1997de,Heinzl:2000ht,Brodsky:2003gk} is an approach to quantum field theory that replaces constant-time hypersurfaces with null hyperplanes. In this paper, we consider the analogous quantization on arbitrary null hypersurfaces (hereafter, ``horizons'') in curved spacetime. A key advantage of lightfront quantization is that the physical vacuum can be constructed kinematically, regardless of interactions \cite{Leutwyler:1970wn}. This is accomplished by defining the vacuum in terms of the generator of null ``time translations'' along the lightfront. We will perform a similar construction for an arbitrary choice of null ``time'' on a general horizon. As the simplest example, we will discuss de Sitter space, where the Bunch-Davies vacuum \cite{Bunch:1978yq} can be viewed \cite{Gibbons:1977mu} as the vacuum with respect to an affine null time along the cosmological horizon. We will rephrase the latter argument within the lightfront approach, stressing that it extends to interacting theories. We will then show how the natural Fock space on the de Sitter horizon captures the standard spatial momentum modes in the cosmological patch.

Physically, null hypersurfaces act as causal boundaries between spacetime regions. In particular, a pair of intersecting horizons divides spacetime into quadrants, of which the two spacelike-separated ones contain the evolution of two ``halves of space''. The entanglement between such regions is an important subject in quantum field theory, with implications for black hole thermodynamics \cite{Bombelli:1986rw,Bianchi:2012ev} and renormalization flows \cite{Myers:2010xs,Casini:2011kv}. It is most often described in terms of Hilbert spaces on \emph{spacelike} hypersurfaces that lie in the appropriate spacetime regions. However, a more natural description would be in terms of the horizons themselves. This is the central goal of this paper. Specifically, we consider a horizon divided into halves along a spatial surface (or, equivalently, an intersection with a second horizon). We then present a recipe for reducing a pure state on the horizon into mixed states on its two halves.
In some cases, e.g. de Sitter horizons and null hyperplanes in flat spacetime, these half-horizon states are causally equivalent to states in the two ``halves of space'' (in the flat case, up to data on a single lightray at null infinity). In other cases, e.g. a Schwarzschild horizon, reconstructing the spatial state requires additional boundary data. However, even then, the state on the horizon may capture the relevant entanglement, as in the Hawking-Unruh effect \cite{Unruh:1976db}.

To illustrate our recipe for restricting states to half-horizons, we will use it to derive the Unruh effect: the vacuum state with respect to a null ``time'' $u$ on a horizon is thermal with temperature $1/2\pi$ with respect to the ``time'' $\tau = \ln u$ on the half-horizon $u>0$. Irrespective of our particular derivation, we stress that the universal form of null horizon kinematics allows us to immediately generalize the Unruh effect to arbitrary null ``time'' parameters on arbitrary curved horizons. We will use this fact to obtain the restriction of a global \emph{thermal} state to a half-horizon, with an application to the causal diamonds of a de Sitter observer.

We assume that the null horizons under consideration are free of caustics. On the other hand, we do not require the horizons to be geodesically complete, so they may be truncated before a caustic is reached. We leave the issue of zero modes in lightfront quantization \cite{Maskawa:1975ky,Yamawaki:1997cj} outside the scope of the paper. For simplicity, we pretend that our quantum field theory contains only scalar fields, with a standard kinetic term and arbitrary potential. For interacting theories, the assumption of a standard kinetic term should be taken with caution, even though it is commonplace in the QCD lightfront quantization literature. The associated subtleties will be briefly discussed in an Appendix.

\section{Operator algebra}

The phase space of a field $\varphi$ on a spacelike or null hypersurface $\Sigma$ consists of the field's value and normal derivative, with symplectic form:
\begin{align}
 \Omega(\delta\varphi_1,\delta\varphi_2) = \int_\Sigma d^3x \left(\delta\varphi_1 S^\mu\del_\mu\delta\varphi_2 - (1\leftrightarrow 2) \right) \ , \label{eq:Omega}
\end{align}
where $S^\mu$ is the densitized normal to $\Sigma$. On a null horizon, $S^\mu$ is the area current, pointing along the horizon's constituent lightrays. A key feature of the null case is that the normal $S^\mu$ is also \emph{tangent} to the horizon; therefore, the values of $\varphi$ on $\Sigma$ determine also $S^\mu\del_\mu\varphi$, and thus span the entire phase space. In this case, the symplectic form \eqref{eq:Omega} becomes:
\begin{align}
 \Omega(\delta\varphi_1,\delta\varphi_2) = \int_\Sigma d^2x\, du \left(\delta\varphi_1 \sqrt{\gamma(u,x)} \del_u\delta\varphi_2 - (1\leftrightarrow 2) \right) \ . \label{eq:Omega_null_before}
\end{align}
Here, $x$ are 2d coordinates labeling the lightrays, $u$ is a null coordinate along each ray,  and $\sqrt{\gamma}$ is the area density of the 2d metric in the $x$ directions. If we now define a rescaled field $\hat \phi$ by:
\begin{align}
 \hat\phi(u,x) \equiv \sqrt[4]{\gamma(u,x)}\,\hat\varphi(u,x) \ , \label{eq:area_rescale}
\end{align}
the symplectic form \eqref{eq:Omega_null_before} becomes:

\begin{align}
 \Omega(\delta\phi_1,\delta\phi_2) = \int_\Sigma d^2x\, du \left(\delta\phi_1  \del_u\delta\phi_2 - (1\leftrightarrow 2) \right) \ . \label{eq:Omega_null}
\end{align}

On horizons where the metric is constant in $u$, the rescaling \eqref{eq:area_rescale} becomes trivial; this case was studied in \cite{Moretti:2004mc}. In general, the rescaling is important as it absorbs the dependence on the metric into the definition of the field $\phi,$ thus rendering the symplectic form \eqref{eq:Omega_null} independent of $\gamma.$  Since the symplectic form, as a functional of the field variations, is independent of the metric, we can import some well-known flat results. In particular, the commutators, obtained by quantizing the Poisson brackets found by inverting the symplectic form \eqref{eq:Omega_null}, can be written as:
\begin{align}
 \left[\hat\phi(u,x), \hat\phi(u',x')\right] = \frac{i}{4}\,\delta^{(2)}(x,x')\sign(u'-u) \ , \label{eq:commutator}
\end{align}
Note that, since they are causally separated, fields on the same lightray do not commute. The expressions \eqref{eq:area_rescale}-\eqref{eq:commutator} (with additional factors) have appeared in the Poisson brackets \cite{Reisenberger:2007ku} for null initial data in General Relativity.

We define creation and annihilation operators by Fourier-transforming $\phi(u,x)$ with respect to the null ``time'' $u$:
\begin{align}
 \hat a(\omega,x) &= \sqrt{2\omega}\int_{-\infty}^\infty du\, e^{i\omega u} \hat\phi(u,x) \ ; \label{eq:a} \\
 \hat a^\dagger(\omega,x) &= \sqrt{2\omega}\int_{-\infty}^\infty du\, e^{-i\omega u} \hat\phi(u,x) \ . \label{eq:a_dagger}
\end{align}
Using \eqref{eq:commutator}, we see that these satisfy the appropriate commutation relations:
\begin{align}
 \begin{split}
   &\left[\hat a(\omega,x), \hat a^\dagger(\omega',x')\right] = 
 2\pi\delta(\omega-\omega')\,\delta^{(2)}(x,x') \ ; \\
   &\left[\hat a(\omega,x), \hat a(\omega',x')\right] = \left[\hat a^\dagger(\omega,x), \hat a^\dagger(\omega',x')\right] = 0 \ .
 \end{split} \label{eq:commutator_a}
\end{align}

Equations \eqref{eq:commutator} and \eqref{eq:commutator_a} giving the commutators of the field $\phi$ and its Fourier modes are the same as one would obtain for $\varphi$ if the horizon was flat. So, while $\phi$ has simple commutation relations, the corresponding relations for $\varphi$ will in general be more complicated. 

The operators \eqref{eq:a}-\eqref{eq:a_dagger} can be used in the standard way to construct e.g. vacuum or thermal states with respect to the ``time translation'' generator $i\del_u$. All of the above is independent of the field's mass and dynamics as encoded in its potential, up to issues with loop corrections that will be discussed in the Appendix.

\section{De Sitter horizon}

As an example, consider a cosmological horizon in de Sitter space. We define de Sitter space as the hyperboloid $v_\mu v^\mu = 1$ within $\mathbb{R}^{1,4}$, invariant under the isometry group $O(4,1)$. The horizon is a 2-sphere of lightrays defined by $\ell_\mu v^\mu = 0$, where $\ell^\mu = (1,1,\vec 0)$ is a null vector in $\mathbb{R}^{1,4}$. The horizon's points can be coordinatized in $\mathbb{R}^{1,4}$ as:
\begin{align}
 v^\mu = (u,u,\vec n) \ . \label{eq:horizon}
\end{align}
Here, the unit 3d vector $\vec n$ plays the role of the lightray label $x$, while $u$ is an affine null time along the rays.

The horizon creation and annihilation operators \eqref{eq:a}-\eqref{eq:a_dagger} have a simple meaning in terms of the Poincare coordinates $(\eta,\vec y)$, which span the cosmological patch to the horizon's future. These are related to the 4+1d radius-vector $v^\mu$ through:
\begin{align}
 v^\mu = -\frac{1}{\eta}\left(\frac{y^2 - \eta^2 + 1}{2}, \frac{y^2 - \eta^2 - 1}{2}, \vec y \right) \ ; \ \eta < 0 \ . \label{eq:Poincare}
\end{align}
The metric is given by:
\begin{align}
 ds^2 = dv_\mu dv^\mu = \frac{1}{\eta^2}(-d\eta^2 + dy^2) \ . \label{eq:metric}
\end{align}
For momentum modes with respect to $\vec y$, the ``infinite past'' $\eta\rightarrow-\infty$ is a UV limit, due to the warp factor in \eqref{eq:metric}. Suppose now that our field theory is well-defined in the UV, by means of a conformal fixed point. Then, although the metric is only \emph{conformally} flat, one can define a Minkowski vacuum at $\eta\rightarrow-\infty$ (noting that any curvature corrections from the conformal transformation are irrelevant in the UV limit). This will be the Bunch-Davies vacuum of the full theory in de Sitter space. 

Now, the horizon \eqref{eq:horizon} can be expressed in the Poincare coordinates \eqref{eq:Poincare} as a particular form of the $\eta\rightarrow-\infty$ limit:
\begin{align}
 \vec y = (-\eta + u)\vec n \quad ; \quad \eta\rightarrow-\infty \ .
\end{align}
In this limit, the time translation $\del_\eta$ becomes the null translation $\del_u$. We conclude that the vacuum annihilated by the horizon operators \eqref{eq:a} is the Minkowski vacuum at $\eta\rightarrow-\infty$, i.e. the Bunch-Davies vacuum.

Note further that spatial translations $\vec y \rightarrow \vec y + \delta\vec y$ of the Poincare coordinates act on the horizon as an $\vec n$-dependent shift $u\rightarrow u + \vec n\cdot\delta\vec y$ along the lightrays. From here, it's easy to see that the creation operators $\hat a^\dagger(\omega,\vec n)$ from \eqref{eq:a_dagger} \emph{create particles with spatial momentum $\vec p = \omega\vec n$ in Poincare coordinates}. This relates the horizon Fock space to the standard cosmological basis of comoving momenta.

\section{Universality of the kinematics and the Unruh effect}

The commutators \eqref{eq:commutator}-\eqref{eq:commutator_a} hold for any null horizon, with $u$ any monotonous null coordinate; $u$ need not be affine, and the range $-\infty<u<\infty$ need not be geodesically complete. For instance, the same relations hold for a null coordinate $\tau = |\ln|u||$ that spans the half-horizon $u>0$ or $u<0$. 

This universality allows us to translate results from any particular horizon and null coordinate to any other. For example, the Unruh effect in flat spacetime states that the vacuum with respect to the null time $u = t+z$ on the horizon $t=z$ is thermal with temperature $1/2\pi$ with respect to the Rindler time $\tau = \ln u$ when restricted to the half-horizon $u>0$. The universality of horizon kinematics then implies that the same is true for the vacuum with respect to any null ``time'' on any null horizon. The Hawking-Unruh effects for de Sitter space and stationary black holes follow as special cases.

For a less standard application, let us find the restriction to $\tau>0$ of a \emph{thermal} state with temperature $T$ with respect to a null time $\tau$. First, we rescale the temperature to $1/2\pi$ through $\tau\rightarrow 2\pi T\tau$. We can now imagine that the thermal state arose from a restriction to $u>0$ of the vacuum with respect to $u=e^{2\pi T\tau}$ (even though $\tau$ may be geodesically complete in the specific spacetime in question). The vacuum with respect to $u$ is also vacuum with respect to $u-1$, and so its restriction to $u>1$, i.e. the restriction of our original thermal state to $\tau>0$, is thermal with temperature $1/2\pi$ with respect to the null coordinate $\ln(u-1) = \ln\left(e^{2\pi T\tau} - 1\right)$.

As a concrete example, consider the de Sitter horizon and $u$ coordinate from the previous section. There, the vacuum with respect to $u$ is the Bunch-Davies vacuum in global de Sitter space. Its restriction to $u>0$ is the thermal state in the causal diamond of the observer with worldline $v^\mu(t) = (\sinh t,\cosh t,\vec 0)$; it has temperature $1/2\pi$ with respect to the static-patch time $t$, or, equivalently, the null time $\tau = \ln u$. The restriction of this thermal state to $\tau>0$ is then thermal with temperature $1/2\pi$ with respect to the null time $\ln(e^\tau - 1)$. Physically, this is the state in the causal diamond of the future half $t>0$ of the observer's worldline.

\section{Wavefunctionals and their restriction to a half-horizon}

Returning now to general null hypersurfaces, consider a horizon divided in half along a spatial section $S$. Each of the horizon's lightrays is divided in half at its intersection with $S$. As an auxiliary structure, we fix some orientation-reversing map between the past and future halves of each lightray. This can always be encoded by choosing a null coordinate $u$ such that $S$ is at $u=0$, and the map between the half-horizons is $u\rightarrow -u$.

To write wavefunctions over fields on the horizon, we should choose a maximal commuting set of field quantities. If the null coordinate range $-\infty<u<\infty$ isn't geodesically complete, our set should also commute with the fields along the lightrays outside this range. It is clear from \eqref{eq:commutator} that these properties are satisfied by the $u$-odd components of the field:
\begin{align}
 \phi_-(u,x) \equiv \frac{1}{2}\left(\phi(u,x) - \phi(-u,x) \right) \quad \text{for} \quad u>0 \ ,
\end{align}
which form a maximal commuting set if we neglect zero modes. As we will see, neglecting zero modes will not affect our main results. Thus, we can represent a pure state on the horizon as a wavefunctional $\Psi[\phi_-(u,x)]$. The following result then specifies how to restrict such a state $|\Psi\rangle$ onto the half-horizon $u>0$.
\begin{theorem} \label{thm:restriction}
 Consider a pure state on the horizon, defined by a wavefunction $\Psi[\phi_-(u,x)]$ over the $u$-odd field components. The state's restriction to the $u>0$ half-horizon can be found in three steps:
 \begin{enumerate}
  \item Replacing $\phi_-(u,x)\rightarrow\phi(u,x)$ for $u>0$, reinterpret $\Psi[\phi_-(u,x)]$ as a functional $\Psi[\phi(u,x);u>0]$ over the half-horizon's phase space.
  \item Using the Wigner-Weyl transform \cite{Wigner}, translate this functional into an operator $\hat\Psi$ on the half-horizon's Hilbert space.
  \item The density operator for the state on the half-horizon is then given by $\hat\rho = \hat\Psi\hat\Psi^\dagger$.
 \end{enumerate}
\end{theorem}
Step 2 is essentially a change of basis, while step 3 contains the actual tracing out of the state at $u<0$.

This theorem provides an efficient way of tracing out the degrees of freedom in a half-horizon, thus giving the density matrix associated with the state on the other half-horizon. This restriction recipe applies to arbitrary states and, to the best of our knowledge, is new even in the flat case.

To prove the Theorem, we follow Unruh \cite{Unruh:1976db} in considering Fourier modes with respect to the null time $\tau = |\ln|u||$ within each half-horizon $u\gtrless 0$. A pure state on each half-horizon can be defined as a wavefunctional over the $\tau$-odd field components, which commute both among themselves and with their counterparts on the other half-horizon. Thus, the task of restricting a state to a half-horizon boils down to transforming from the $u$-odd basis to the $\tau$-odd one. 

The $\tau$-odd fields on the $u>0$ and $u<0$ half-horizons are spanned respectively by the Fourier modes:
\begin{align}
 \begin{split}
   B_1(\Omega,x) &= 2\sqrt{\Omega}\int_{-\infty}^\infty d\tau \sin(\Omega\tau)\, \phi(e^\tau,x) \ ; \\
   B_2(\Omega,x) &= 2\sqrt{\Omega}\int_{-\infty}^\infty d\tau \sin(\Omega\tau)\, \phi(-e^{-\tau},x) \ ,
 \end{split} 
\end{align}
while the $\tau$-even fields are spanned by:
\begin{align}
 \begin{split}
   A_1(\Omega,x) &= 2\sqrt{\Omega}\int_{-\infty}^\infty d\tau \cos(\Omega\tau)\, \phi(e^\tau,x) \ ; \\
   A_2(\Omega,x) &= 2\sqrt{\Omega}\int_{-\infty}^\infty d\tau \cos(\Omega\tau)\, \phi(-e^{-\tau},x) \ ,
 \end{split} 
\end{align}

The $u$-odd fields on the full horizon are spanned by:
\begin{align}
 \begin{split}
   A(\Omega,x) &= 2\sqrt{\Omega}\int_{-\infty}^\infty d\tau \cos(\Omega\tau)\, \phi_-(e^\tau,x) \ ; \\
   B(\Omega,x) &= 2\sqrt{\Omega}\int_{-\infty}^\infty d\tau \sin(\Omega\tau)\, \phi_-(e^\tau,x) \ ,
 \end{split} \label{eq:A_B}
\end{align}
where $B = (B_1 + B_2)/2$. Similarly to \eqref{eq:commutator_a}, we see that $A$ is canonically conjugate to $B_2 - B_1$. Thus, the transformation from the $u$-odd basis to the $\tau$-odd one takes the form:
\begin{align}
 \tilde\Psi[B_1,B_2] = \int\! \mathcal{D}\!A\, e^{iA\cdot(B_2 - B_1)}\, \Psi\!\left[A,\textstyle\frac{1}{2}(B_1 + B_2)\right] \ , \label{eq:WW}
\end{align}
where the scalar products $A\cdot B_{1,2}$ stand for:
\begin{align}
 A\cdot B_{1,2} \equiv \int d^2x \int_0^\infty \frac{d\Omega}{2\pi}\,A(\Omega,x)B_{1,2}(\Omega,x) \ . 
\end{align}
The density matrix elements of the state restricted to the $u>0$ horizon now read:
\begin{align}
 \rho[B_1,B'_1] = \int \mathcal{D}B_2\, \tilde\Psi[B_1,B_2]\,\tilde\Psi^*[B'_1,B_2] \ . \label{eq:trace}
\end{align}

Finally, notice that if we substitute $\phi_- \rightarrow \phi$ in \eqref{eq:A_B}, then $A$ and $B = (B_1+B_2)/2$ become canonical conjugates, spanning the phase space at $u>0$. Eq. \eqref{eq:WW} can now be recognized as a Wigner-Weyl transform between phase-space functions and operator matrix elements, while eq. \eqref{eq:trace} becomes a matrix product. These observations bring the state-restriction recipe \eqref{eq:WW}-\eqref{eq:trace} into the form given in the Theorem.

Moreover, notice that the procedure described in the Theorem is consistent with neglecting zero modes, since the non-zero $\tau$ modes on the half horizons do not depend on the zero $u$ modes of the field on the whole horizon. This can be seen by shifting the field on the full horizon by a constant. If $\phi \rightarrow \phi + c,$ then:
\begin{align}
 \begin{split}
 A_{1,2}(\Omega,x) \rightarrow A_{1,2}(\Omega,x) + 2 c \sqrt{\Omega}\int_{-\infty}^\infty d\tau \cos(\Omega\tau), \, \\
 B_{1,2}(\Omega,x) \rightarrow B_{1,2}(\Omega,x) + 2 c \sqrt{\Omega}\int_{-\infty}^\infty d\tau \sin(\Omega\tau). \,
 \end{split} \label{eq:ABchange}
\end{align}
So, after regularizing these integrals, unless $\Omega=0,$ both $A$ and $B$ modes remain unchanged. 

\section{Deriving the Unruh effect}

In this section, we illustrate our state-restriction recipe by applying it to the vacuum state on a null horizon. This will yield the Unruh effect, together with its generalizations discussed above. Horizon-based approaches to the Unruh effect are of course not new; see e.g. \cite{Sewell:1982zz}.

We begin by expressing the vacuum wavefunctional $\Psi_0[\phi_- (u,x)]$ in the basis of $u$-odd fields. First, define the sine and cosine transforms of $\phi(u,x)$ as:
\begin{align}
 \begin{split}
   f(\omega,x) &= 2\sqrt{\omega}\int_{-\infty}^\infty du\, \cos(\omega u)\, \phi(u,x) \ ; \\
   g(\omega,x) &= 2\sqrt{\omega}\int_{-\infty}^\infty du\, \sin(\omega u)\, \phi(u,x) \ ,
 \end{split}
\end{align}
satisfying the commutation relations:
\begin{align}
 \begin{split}
   &\left[\hat f(\omega,x), \hat g(\omega',x')\right] = 2\pi i\,\delta(\omega-\omega')\,\delta^{(2)}(x,x') \ ; \\
   &\left[\hat f(\omega,x), \hat f(\omega',x')\right] = \left[\hat g(\omega,x), \hat g(\omega',x')\right] = 0 \ .
 \end{split} \label{eq:commutator_fg}
\end{align}
The sine coefficients $g(\omega,x)$ form a basis for the $u$-odd fields. In this basis, the lowering operators \eqref{eq:a} become:
\begin{align}
 \begin{split}
   \hat a(\omega,x) &= \frac{1}{\sqrt{2}}\left(\hat f(\omega,x) + i\hat g(\omega,x)\right) \\
     &= \frac{i}{\sqrt{2}}\left(2\pi \frac{\delta}{\delta g(\omega,x)} + g(\omega,x) \right) \ .
 \end{split}
\end{align}
From now on, we omit the spatial coordinates $x$, since they are simply carried around or integrated over. The vacuum wavefunction defined by $\hat a(\omega)\Psi_0 = 0$ reads:
\begin{align}
 \Psi_0 \sim \exp \left( -\frac{1}{2} \int_{0}^\infty\frac{d\omega}{2\pi}\, g(\omega)^2 \right) \ ,
\end{align}
or, explicitly in terms of $\phi_-(u)$:
\begin{align}
 \Psi_0 \sim \exp \left(\frac{8}{\pi}\int_0^\infty\!\! \int_0^\infty \! \frac{uu' dudu'}{(u^2-u'^2)^2}\, \phi_-(u) \phi_-(u') \right) \ . \label{eq:Psi_phi_-}
\end{align}

We now follow the Theorem's recipe to find the density matrix of the state restricted to the $u>0$ half-horizon. Thus, we substitute $\phi_-(u)\rightarrow\phi(u)$ in \eqref{eq:Psi_phi_-}. We then rewrite the resulting functional in terms of harmonic oscillators with respect to the null time $\tau = \ln u$:
\begin{align}
 \begin{split}
   a_\tau(\Omega) &= \sqrt{2\Omega}\int_{-\infty}^\infty d\tau\, e^{i\Omega\tau} \phi(e^\tau) \ ; \\
   a^*_\tau(\Omega) &= \sqrt{2\Omega}\int_{-\infty}^\infty d\tau\, e^{-i\Omega\tau} \phi(e^\tau) \ ,
 \end{split}
\end{align}
where the subscript $\tau$ is to distinguish these from the oscillators \eqref{eq:a}-\eqref{eq:a_dagger} with respect to $u$. The wavefunction \eqref{eq:Psi_phi_-} becomes:
\begin{align}
 \Psi_0 \sim \exp \left(-2 \int_0^{\infty} \frac{d\Omega}{2\pi} \coth\left(\frac{\pi\Omega}{2}\right) a_\tau^*(\Omega) a_\tau(\Omega) \right) \ . \label{eq:Psi_a}
\end{align}
We now use the well-known Wigner-Weyl transform of a Gaussian (neglecting normalization coefficients):
\begin{align}
 \exp\left(-\beta\hat a^\dagger\hat a \right) \ \longleftrightarrow \ \exp\left(-2\tanh\left(\frac{\beta}{2}\right) a^* a \right) \ .
\end{align}
This turns \eqref{eq:Psi_a} into the operator:
\begin{align}
 \begin{split}
   \hat\Psi_0 &\sim \exp \left(-\pi \int_0^\infty \frac{d\Omega}{2\pi}\,(\Omega + i)\, \hat a^\dag_\tau(\Omega) \hat a_\tau(\Omega) \right) \\
    &= (-1)^{\hat N}e^{-\pi\hat H} \ ,
 \end{split} \label{eq:Psi_hat}
\end{align}
where $\hat H$ is the Hamiltonian for $\tau$ translations, while $\hat N$ is the number operator for quanta created by the $\hat a^\dag_\tau$'s. Squaring the operator \eqref{eq:Psi_hat}, we obtain the density matrix for the state on the half-horizon $u>0$:
\begin{align}
 \hat\rho = \hat\Psi_0 \hat\Psi_0^\dagger \sim e^{-2\pi\hat H} \ .
\end{align}
As expected, this is the thermal state with temperature $1/2\pi$.

\section*{Acknowledgements}		

We are grateful to Laurent Freidel, Rob Myers and Vasudev Shyam for discussions, as well as to Bill Unruh and Aron Wall for email exchanges, and to an anonymous referee for helpful comments. Research at Perimeter Institute is supported by the Government of Canada through Industry Canada and by the Province of Ontario through the Ministry of Research \& Innovation. YN also acknowledges support of funding from NSERC Discovery grants.

\appendix
\section{Loop corrections and the commutator in interacting theories}

In the bulk of this paper, we've been cavalier about the transition from the classical symplectic form \eqref{eq:Omega_null} to the quantum commutator \eqref{eq:commutator}. For free fields, this step is straightforward. There is similarly no problem for interacting fields at tree level. However, greater care is needed once loop corrections are taken into account. Indeed, while the symplectic form \eqref{eq:Omega_null} assumes a Lagrangian with a standard kinetic term, the Lagrangian becomes a scale-dependent object. We will now discuss briefly the possible scenarios, drawing on the discussion in \cite{Wall:2011hj}. The issue is not tied to our consideration of curved horizons, and is also relevant for e.g. the flat-space lightfront quantization of QCD.

One possibility is to define the theory with a UV cutoff. Then our field operators and the Lagrangian from which we derive \eqref{eq:Omega_null}-\eqref{eq:commutator} can simply live at the cutoff. For our purposes, the cutoff would have to respect the fact that the horizon is foliated by lightrays. For instance, one could replace the 2d set of lightrays by a discrete 2d lattice, while keeping each ray null and continuous.

Suppose now that we're interested instead in a continuum theory. Then our first observation is that the commutator \eqref{eq:commutator} should be treated as a UV object, since the only causal separations on the horizon are of zero length. We are therefore interested in the Lagrangian at an energy scale $\mu\rightarrow\infty$. At a bare minimum, this requires our theory to be renormalizable with a (conformal) UV fixed point. However, this may not be enough. The reason is that even with a UV fixed point, there may be an infinite renormalization of the field between a finite scale $\mu$ accessible to experiments and the $\mu\rightarrow\infty$ limit. This translates into an infinite renormalization of the Lagrangian's kinetic term, calling into question the whole reasoning leading to the commutator \eqref{eq:commutator}. Whether or not this problem arises depends on the weakness of interactions in the UV. 

Assume the standard behavior $\sim g^2 d(\ln\mu)$ for the field strength renormalization, where $g$ is the dimensionless coupling. If the coupling remains finite at the UV fixed point, the field renormalization is infinite, diverging as $\ln\mu$; this is a manifestation of the field's anomalous dimension. If the UV coupling vanishes, i.e. for theories with a free fixed point, the behavior will depend on how quickly $g$ goes to zero as a function of $\mu$. For super-renormalizable theories, where $g$ decays polynomially, the field renormalization is finite, and lightfront quantization is clearly valid. On the other hand, for a theory like QCD, with the UV behavior $g^2\sim 1/\ln\mu$, the field renormalization is infinite, diverging as $\sim\ln\ln\mu$. Thus, without discounting its potential utility, the theoretical foundations of lightfront QCD remain unclear.

\end{document}